\documentclass[useAMS,usenatbib]{mn2e}
\usepackage{epsfig}
\usepackage{amssymb}
\usepackage{amsmath}
\usepackage{lscape}
\usepackage{longtable}
\usepackage[normal]{caption}
\usepackage{appendix}
\usepackage{bm}
\usepackage{graphicx}

\title[What determines the shape of the local ($z<0.1$) infrared galaxy luminosity function?]{What determines the shape of the local ($z<0.1$) infrared galaxy luminosity function?}

\author[M.~Symeonidis and M. J. ~Page] 
{\parbox{\textwidth}{\raggedright
M.~Symeonidis,$^{1}$\thanks{E-mail: \texttt{m.symeonidis@ucl.ac.uk}}
and M. J. ~Page,$^{1}$ }\vspace{0.4cm}\\
\parbox{\textwidth}{\raggedright $^{1}$ Mullard Space Science
  Laboratory, University College London, Holmbury St. Mary, Dorking,
  Surrey RH5 6NT, UK}}

\begin{document}

\date{Accepted  Received; in original form}

\pagerange{\pageref{firstpage}--\pageref{lastpage}} \pubyear{2014}

\maketitle

\label{firstpage}

\begin{abstract}
We investigate what shapes the infrared luminosity function of local galaxies by comparing it to the local infrared AGN luminosity function. The former corresponds to emission from dust heated by stars and AGN, whereas the latter includes emission from AGN-heated dust only. Our results show that infrared emission from AGN starts mixing into the galaxy luminosity function in the luminous infrared galaxy (LIRG) regime and becomes significant in the ultraluminous infrared galaxy (ULIRG) regime, with the luminosity above which local ULIRGs become AGN-dominated being in the log\,$L_{\rm IR}/\rm L_{\odot}\sim12.2$--$12.7$ range. We propose that as a result of the AGN contribution, the infrared galaxy luminosity function has a flatter high luminosity slope than UV/optical galaxy luminosity functions. Furthermore, we note that the increased AGN contribution as a function of $L_{\rm IR}$ is reflected in the average dust temperature ($T_{\rm dust}$) of local galaxies, and may be responsible for the local $L_{\rm IR}$-$T_{\rm dust}$ relation. However, although our results show that AGN play a central role in defining the properties of local ULIRGs, we find that the dominant power source in the local ULIRG population is star-formation.

\end{abstract}

\begin{keywords}
 galaxies: luminosity function, mass function 
 galaxies: star formation 
 infrared: galaxies
 (galaxies:) quasars: general 
 \end{keywords}

\section{Introduction}
\label{sec:introduction}

Although infrared radiation was first associated with individual galaxies in the late 1960s (e.g. Johnson 1966\nocite{Johnson66}; Low $\&$ Tucker 1968\nocite{LT68}; Kleinmann $\&$ Low 1970\nocite{KL70}), it was the infrared all-sky survey with \textit{IRAS} (Neugebauer et al. 1984\nocite{Neugebauer84}), later repeated with \textit{AKARI} (Murakami et al. 2007\nocite{Murakami07}), that revealed a large number of dust-enshrouded galaxies in the local ($z<0.1$) Universe, with total infrared luminosities ($L_{\rm IR}$, 8---1000\,$\mu$m) up to $10^{13}\,\rm L_{\odot}$ (e.g. Houck et al. 1984\nocite{Houck84}; 1985\nocite{Houck85}; Soifer et al. 1984a\nocite{Soifer84a}, 1984b\nocite{Soifer84b}). 

The infrared (IR) luminosity function (LF) of local ($z<0.1$) galaxies, first examined using \textit{IRAS} data (Soifer et al. 1986\nocite{Soifer86}; 1987\nocite{Soifer87}), was seen to diverge from the Schechter function shape (Schechter 1976\nocite{Schechter76}) that characterises the optical LF of local galaxies: it displays a shallower drop off at the high luminosity end (Soifer et al. 1987\nocite{Soifer87}). As a result, it is normally fit with a double power-law (e.g. Lawrence et al. 1986\nocite{Lawrence86}; Soifer et al. 1987\nocite{Soifer87}; Sanders et al. 2003\nocite{Sanders03}) or a combination of power law for $L<L_{\star}$ and a Gaussian in log\,$L$ for $L>L_{\star}$ (e.g. Saunders et al. 1990\nocite{Saunders90}). 

The high luminosity end ($L>L_{\star}$) of the local IR LF is made up of ultraluminous infrared galaxies (ULIRGs), defined as galaxies with $L_{\rm IR}$=$10^{12}-10^{13} \rm L_{\odot}$ (e.g. Sanders $\&$ Mirabel 1996\nocite{SM96}; Genzel et al. 1998\nocite{Genzel98}). ULIRGs are characterized by warm average dust temperatures ($>30\,K$; e.g. Soifer et al. 1984b\nocite{Soifer84b}; Klaas et al. 1997\nocite{Klaas97}; Clements et al. 2010\nocite{CDE10}) and strong silicate absorption in their mid-infrared continua (e.g. Armus et al. 2007\nocite{Armus07}). The AGN incidence is high in ULIRGs, with the majority of them residing in the AGN region in optical line ratio diagrams (e.g. Sanders et al. 1988a\nocite{Sanders88a}) and many hosting buried AGN discovered through mid-infrared spectroscopy and X-ray observations (e.g. Imanishi et al. 2007\nocite{Imanishi07}; 2008\nocite{Imanishi08}; 2010\nocite{IMN10}; Armus et al. 2006\nocite{Armus06}; Oyabu et al. 2011\nocite{Oyabu11}). The primary energy source in ULIRGs has thus always been a topic of much contention. Since the `Great Debate' of 1999 (Sanders 1999\nocite{Sanders99}; Joseph 1999\nocite{Joseph99}) where the case was made for and against AGN as the primary energy source in local ULIRGs, new data have not converged to an answer, and still the only indisputable fact is the composite AGN/starburst nature of these sources (e.g. Gregorich et al. 1995\nocite{Gregorich95}; Genzel et al. 1998\nocite{Genzel98}; Soifer et al. 2000\nocite{Soifer00}; Klaas et al. 2001\nocite{Klaas01}; Davies et al. 2002\nocite{DBW02}; Franceschini et al. 2003\nocite{Franceschini03b}).

Here we examine what determines the shape of the IR LF of local ($z<0.1$) galaxies, focusing on its high luminosity tail, with the additional aim of gaining insight into the nature of ULIRGs. The letter is laid out as follows: in sections \ref{sec:method} and \ref{sec:results} we describe our method and results. The discussion and conclusions are presented in sections \ref{sec:discussion} and \ref{sec:conclusions}. Throughout, we adopt a concordance cosmology of H$_0$=70\,km\,s$^{-1}$Mpc$^{-1}$, $\Omega_{\rm M}$=1-$\Omega_{\rm \Lambda}$=0.3.

\begin{figure*}
\epsfig{file=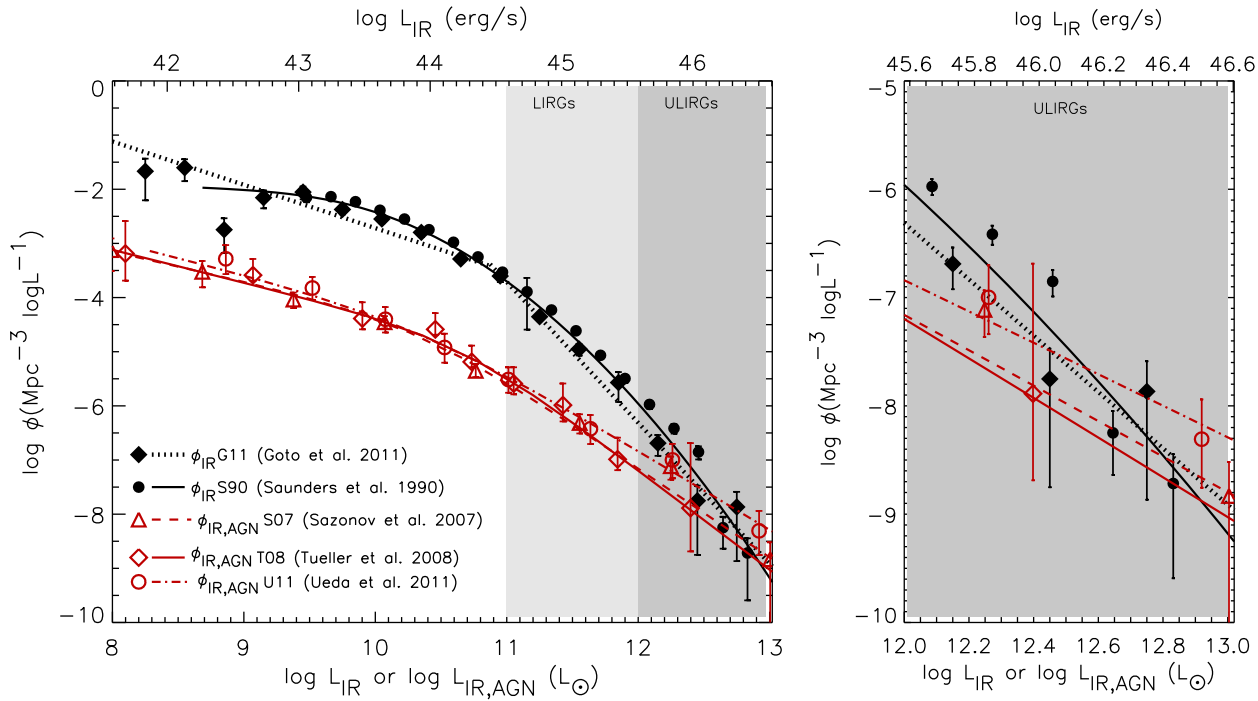,width=0.99\linewidth} \\
\caption{\textit{Left panel}: The two realisations of $\phi_{\rm IR}$ (in black) and the three realisations of $\phi_{\rm IR, AGN}$ (in red). The data and the model fits are shown in all cases. \textit{Right panel}: Same as left but zoomed in to the ULIRG regime.} 
\label{fig:fig1}
\end{figure*}

\section{Method}
\label{sec:method}

\begin{figure}
\epsfig{file=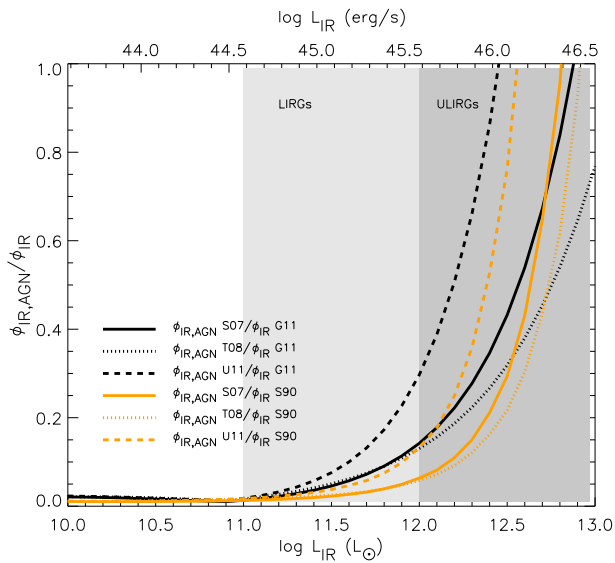,width=0.99\linewidth} \\
\caption{Plot of the $\phi_{\rm IR, AGN}$ to $\phi_{\rm IR}$ ratio --- a simple estimate of the fraction of AGN-dominated sources as a function of $L_{\rm IR}$. There are 3 realisations of $\phi_{\rm IR, AGN}$ and 2 of $\phi_{\rm IR}$, so 6 realisations of this ratio. The key is S07: Sazonov et al. (2007); G11: Goto et al. (2011); T08: Tueller et al. 2008; U11: Ueda et al. (2011); S90: Saunders et al. (1990).} 
\label{fig:fig2}
\end{figure}

For our investigation we compare the IR LF of galaxies to that of AGN, similar to the investigation presented in Symeonidis $\&$ Page (2018\nocite{SP18}; hereafter SP18) for $1<z<2$ hyperluminous infrared galaxies (HyLIRGs). 

The best estimate of the local IR LF (hereafter $\phi_{\rm IR}$) comes from the \textit{IRAS} and \textit{AKARI} all-sky surveys. We use the local $\phi_{\rm IR}$ as a function of $L_{\rm IR}$ from Saunders et al. (1990; hereafter S90) built with \textit{IRAS} data\footnote{From S90 we chose their default $\phi_{\rm IR}$, calculated with flow model V3, $H_{0}$=66\,km/s/Mpc, $\Omega$=1 and $\Lambda$=0 based on the S1-S7 samples (see S90 for more details)} and from Goto et al. (2011\nocite{Goto11a}; hereafter G11) built with \textit{AKARI} data. The main difference in the S90 and G11 parametric models is that the former is a combination of a power law slope for $L<L_{\star}$ and a Gaussian in log\,$L$ for $L>L_{\star}$, whereas the latter is described as a double power-law. The S90 $\phi_{\rm IR}$ was originally built as a function of 60$\mu$m monochromatic luminosity rather than $L_{\rm IR}$, so for our purposes we convert to the latter as follows: using the Chary $\&$ Elbaz (2001) SED library, we estimate the linear relation between 60$\mu$m monochromatic luminosity and $L_{\rm IR}$ (8--1000$\mu$m) for log\,[$\nu L_{\nu, 60}/L_{\odot}$]$<$10.5 and for log\,[$\nu L_{\nu, 60}/L_{\odot}$]$>$10.5 separately, yielding log\,[$L_{\rm IR}]=0.41+ 0.98\, \rm log [\nu L_{\nu, 60}]$ for the former and log\,[$L_{\rm IR}]=1.23+ 0.91\, \rm log [\nu L_{\nu, 60}]$ for the latter. We subsequently use these scaling relations to convert log\,[$\nu L_{\nu, 60}$] to $L_{\rm IR}$ and correct the space densities for the difference in bin size. As the focus of our work is the behaviour of the luminosity functions in the ULIRG regime, we test the CE01 library against a sample of local ULIRGs with the most most up-to-date measurements of $L_{\rm IR}$ which use \textit{Herschel} data (Clements et al. 2018\nocite{Clements18}). We find that the average $L_{\rm IR}/ \nu L_{\nu, 60}$ ratio of local ULIRGs is consistent with the average $L_{\rm IR}/ \nu L_{\nu, 60}$ ratio of the CE01 library in that luminosity range.

For the AGN luminosity function (LF), we use three derivations of the absorption-corrected AGN LF: the 17---60\,keV X-ray LF from Sazonov et al. (2007\nocite{Sazonov07}; hereafter S07) using data from the \textit{INTEGRAL} all-sky survey, the Tueller et al. (2008\nocite{Tueller08}; hereafter T08) 14-195\,keV LF using data from Swift's Burst Alert Telescope (BAT) and the 2--10\,kev LF from Ueda et al. (2011; hereafter U11) using data from the Monitor of All-sky X-ray Image (MAXI) mission on the International Space Station (Matsuoka  et  al.  2009). All three LFs exclude blazars. For our purposes, the abscissae of all LFs are converted to intrinsic infrared AGN luminosity in the 8--1000$\mu$m range ($L_{\rm IR, AGN}$) as follows: first the T08 and S07 LFs are converted to the 2-10keV X-ray band using a photon index $\Gamma$=2, chosen as it is mid-way in the range of measured values of 1.5-2.5 (e.g. Nandra $\&$ Pounds 1994\nocite{NP94}; Reeves $\&$ Turner 2000\nocite{RT00}; Page et al. 2005\nocite{Page05}), but also consistent with the median spectral index reported in Beckmann et al. (2006b) and T08.  For all three LFs, the 2--10\,keV luminosity is then converted to optical luminosity at 5100$\AA$ ($\nu L_{\nu, 5100}$), adopting the relation from Maiolino et al. (2007\nocite{Maiolino07}) and subsequently to infrared luminosity in the 8--1000$\mu$m range ($L_{\rm IR, AGN}$) using the intrinsic AGN SED of Symeonidis et al. (2016; hereafter S16\nocite{Symeonidis16}). The space densities are also corrected to take into account the different bin sizes from the X-ray to the optical. 

The two realisations of $\phi_{\rm IR}$ and the three realisations of $\phi_{\rm IR, AGN}$ are shown in Fig \ref{fig:fig1}. Note that the T08 $\phi_{\rm IR, AGN}$ consists of sources at $z<0.1$, but the S07 $\phi_{\rm IR, AGN}$ includes an object at z=0.14 and the U11 $\phi_{\rm IR, AGN}$ includes an object with z=0.186, hence they extend to higher luminosities than the T08 $\phi_{\rm IR, AGN}$. $\phi_{\rm IR}$ and $\phi_{\rm IR, AGN}$ are monotonically decreasing functions of $L_{\rm IR}$ and $L_{\rm IR, AGN}$ respectively. $L_{\rm IR, AGN}$ is the intrinsic IR luminosity of the AGN, i.e. IR emission from AGN-heated dust only, whereas $L_{\rm IR}$ includes the total dust-reprocessed emission from stars and AGN. As a result, $\phi_{\rm IR}$ should include all sources that make up $\phi_{\rm IR, AGN}$ and the condition which characterises the two luminosity functions is thus $\phi_{\rm IR} \geq \phi_{\rm IR, AGN}$. This is indeed corroborated observationally, for example many \textit{IRAS} galaxies host X-ray detected AGN (e.g. Franceschini et al. 2003\nocite{Franceschini03}, Teng et al. 2009\nocite{Teng09}; Iwasawa et al. 2011\nocite{Iwasawa11}) and the vast majority of the AGN in the samples we use for the X-ray LFs are also bright \textit{IRAS} galaxies (see also Vasudevan et al. 2010\nocite{Vasudevan10}). There is also a population of Compton thick AGN which do not feature in the X-ray luminosity function but would be contributing to $\phi_{\rm IR, AGN}$; indeed many luminous infrared galaxies not detected in the X-rays are thought to host Compton thick AGN (e.g. Imanishi et al. 2007\nocite{Imanishi07}; Nardini $\&$ Risaliti 2011\nocite{NR11}). To take these sources into account we assume a local Compton thick AGN fraction ($f_{\rm CT}$). Although $f_{\rm CT}$ values reported in the literature range from 20 per cent (e.g. Brightman $\&$ Nandra 2011b\nocite{BN11b}) to 50 per cent (e.g. Maiolino et al. 1998\nocite{Maiolino98}; Guainazzi et al. 2005\nocite{Guainazzi05}), the most recent ones are closer to 30 per cent (e.g. Ricci et al. 2015\nocite{Ricci15}) so, hereafter, we use $f_{\rm CT}$=0.3. We multiply all three realisations of $\phi_{\rm IR, AGN}$ by $\frac{1}{1+f_{\rm CT}}$ to get the final $\phi_{\rm IR, AGN}$.

\section{Results}
\label{sec:results}

Fig \ref{fig:fig1} shows that initially (for $L_{\rm IR}<10^{11}$~L$_{\odot}$), $\phi_{\rm IR}$ exceeds $\phi_{\rm IR, AGN}$ by more than 1\,dex. However, this difference decreases with increasing luminosity because $\phi_{\rm IR}$ declines faster than $\phi_{\rm IR, AGN}$ and at $L_{\rm IR} > 10^{12}$L$_{\odot}$ the two luminosity functions converge. The point of convergence in the parametric models of the luminosity functions takes place in the ULIRG regime, where the space densities of AGN and galaxies are consistent within the errors. Note that although the parametric models of the LFs seem to cross-over (Fig \ref{fig:fig1}), this is simply the effect of extrapolating them. In reality the two LFs never cross over and by definition the condition $\phi_{\rm IR} \geq \phi_{\rm IR, AGN}$ always holds. At the high luminosity end, the number densities of AGN and infrared sources are consistent within the errors, suggesting that $\phi_{\rm IR} = \phi_{\rm IR, AGN}$.

We remind the reader that there is 100 per cent overlap between $\phi_{\rm IR, AGN}$ and $\phi_{\rm IR}$, in the sense that all sources in the former are also part of the latter. The ratio of $\phi_{\rm IR, AGN}$ to $\phi_{\rm IR}$ thus provides a simple estimate of the fraction of AGN-dominated sources as a function of $L_{\rm IR}$ (see also SP18). Interpreting the ratio in this way assumes that the AGN/star formation dominance is a binary process, where galaxies are either entirely AGN-powered or  star formation-powered (i.e. there is no mixing) and $\phi_{\rm IR, AGN}$/$\phi_{\rm IR}$ essentially represents $n_{\rm AGN}/(n_{\rm AGN}+n_{\rm SF}$), where $n_{\rm AGN}$ is the number of AGN powered galaxies and $n_{\rm SF}$ is the number of star formation powered galaxies. Although this is not true at low luminosities, as we approach the high luminosity end of the luminosity function, there is convergence to the condition $L_{\rm IR}$=$L_{\rm IR, AGN}$, i.e. the AGN infrared emission makes up the whole $L_{\rm IR}$. As a result, we expect that the ratio of $\phi_{\rm IR, AGN}$ to $\phi_{\rm IR}$ adequately traces the AGN-dominated fraction of galaxies (at least) in the ULIRG regime. Fig. \ref{fig:fig2} shows the ratio $\phi_{\rm IR, AGN}$ to $\phi_{\rm IR}$ as a function of $L_{\rm IR}$. $\phi_{\rm IR, AGN}$/$\phi_{\rm IR}$ is calculated from the parametric models of the luminosity functions shown in Fig. \ref{fig:fig1}. 

We calculate six realisations of this ratio, since there are three realisations of $\phi_{\rm IR, AGN}$ and two of $\phi_{\rm IR}$ (Fig. \ref{fig:fig2}). The minimum, mean and maximum is shown in Fig. \ref{fig:fig3} (top panel). Note the general trend: the fraction of AGN-dominated sources is small, $<3$ per cent, until the LIRG regime, where it starts increasing and eventually the population becomes AGN dominated. The luminosity above which ULIRGs become AGN-dominated (i.e. $\phi_{\rm IR, AGN}/\phi_{\rm IR}$ $>50\%$) lies somewhere in the range of log\,$L_{\rm IR}/\rm L_{\odot}\sim12.2$ to log\,$L_{\rm IR}/\rm L_{\odot} \sim12.7$.

\section{Discussion}
\label{sec:discussion}

\subsection{The shape of the local infrared luminosity function}
\label{sec:densities}

The work we present here compares the space densities of galaxies and AGN as a function of infrared luminosity. For the first time, we express the AGN LF in terms of total infrared power, allowing the AGN infrared emission to be treated separately to that of its host galaxy, hence enabling insight into the contribution of AGN to galaxies' infrared energy output. We find that $\phi_{\rm IR, AGN}$ and $\phi_{\rm IR}$ are initially offset but converge in the ULIRG regime, where for a given $L_{\rm IR}$, the space densities of galaxies are consistent within the errors, with the space density of AGN. Since $\phi_{\rm IR} \geq \phi_{\rm IR, AGN}$, the convergence of $\phi_{\rm IR}$ and $\phi_{\rm IR, AGN}$ indicates that AGN play a role in shaping the high luminosity end of the local IR LF. A simple measure of the AGN contribution to $L_{\rm IR}$, parametrised by the $\phi_{\rm IR, AGN}$/$\phi_{\rm IR}$ ratio (Fig \ref{fig:fig2}), shows that it is small, $<3$ per cent, until the LIRG regime but increases rapidly thereafter. The increasing AGN contribution is easily accommodated by the AGN incidence rate, the latter rising ahead of the former as a function of $L_{\rm IR}$: G11 shows that in the LIRG regime the AGN incidence rate is 30--70 per cent and in the ULIRG regime the vast majority of sources host AGN (see also Brand et al. 2006\nocite{Brand06}; Nardini $\&$ Risaliti 2011\nocite{NR11}).

The first comparisons between the space densities of galaxies and AGN were reported in the original works on the IR LF by Soifer et al. (1986\nocite{Soifer86}; 1987\nocite{Soifer87}) and Sanders et al. (1988a\nocite{Sanders88a}; 1988b\nocite{Sanders88b}; 1989\nocite{Sanders89}),where the comparison was made in terms of bolometric luminosity. Soifer et al. (1987) noted that from about $L_{\rm IR} \sim 10^{11} \rm L_{\odot}$ upwards the space densities of Seyferts matched those of the \textit{IRAS} sample and Sanders et al. (1988b) found that PG QSOs had similar space densities to warm ULIRGs (defined as those with $f_{\nu}$ (25$\mu$m)/$f_{\nu}$ (60$\mu$m)$>$0.2). These results were part of the motivation for the well-known Sanders et al. (1988a; 1988b) hypothesis that ULIRGs evolve into unobscured QSOs. Although the work we present here does not shed light on whether obscured AGN evolve into unobscured AGN, it indicates that the likely reason for the similarity in space densities between ULIRGs and AGN is that AGN produce a significant fraction of the infrared emission in ULIRGs. 

We propose that AGN alter the high luminosity slope of the local IR LF causing it to be flatter than the traditional Schechter function shape. From $10^{11} \rm L_{\odot}$, where $\phi_{\rm IR, AGN}$/$\phi_{\rm IR}$ starts increasing, the infrared emission from AGN heated dust mixes with that from stellar heated dust, augmenting the galaxies' total infrared luminosity, subsequently shifting them to a higher luminosity bin, hence flattening the LF slope. Eventually AGN take over entirely and at this point the slope of the IR LF is equal to that of the AGN LF. The latter effect is clearly seen at $1<z<2$ (SP18). In the local Universe this is expected to happen somewhere in the log\,[$L_{\rm IR} / \rm L_{\odot}$]=12.4--13.1 range (see Fig. \ref{fig:fig2}). However there are not enough objects to accurately measure the shape of $\phi_{\rm IR}$ at those luminosities, so this effect is missed. Contrary to $\phi_{\rm IR}$ which includes contribution from all AGN irrespective of obscuration or luminosity, the AGN influence is not seen in the local optical/UV galaxy LFs because obscured AGN contribute very little in the UV/optical and the luminous unobscured AGN are readily identified and removed.

\begin{figure}
\epsfig{file=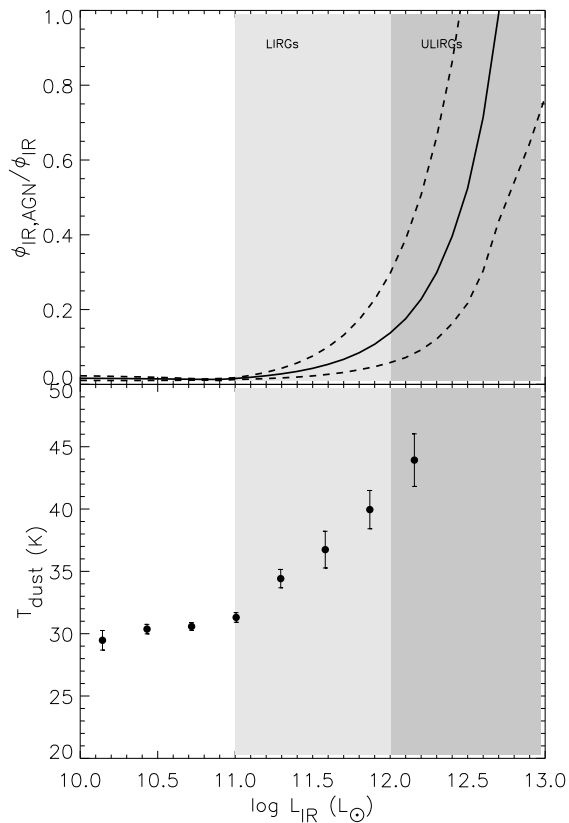,width=0.85\linewidth} 
\caption{Top panel: the minimum, mean and maximum $\phi_{\rm IR, AGN}$/$\phi_{\rm IR}$ ratio. Lower panel: the dust temperature as a function of $L_{\rm IR}$ for local galaxies taken from Symeonidis et al. (2013).}
\label{fig:fig3}
\end{figure}

\subsection{The luminosity--dust temperature relation}
\label{sec:LT}

AGN SEDs are flatter in the mid-IR (e.g. de Grijp et al 1985\nocite{deGrijp85}) than the typical star-forming galaxy SED, because IR emission from AGN is dominated by emission from dust in the torus, which reaches near-sublimation temperatures ($<$1200\,K; e.g. Rodriguez-Ardila $\&$ Mazzalay 2006\nocite{RAM06}) and therefore peaks at 5--20$\mu$m (e.g. Sanders et al. 1988a; 1988b; Sanders $\&$ Mirabel 1996\nocite{SM96}; S16), unlike emission from stellar-heated dust which is characterised by cooler average dust temperatures ($T_{\rm dust}$$\sim$20-40\,K) and a peak at longer wavelengths (60-100$\mu$m), e.g. Symeonidis et al. (2013\nocite{Symeonidis13a}; hereafter S13). Mixing hot dust emission from the AGN with cooler emission from stellar-heated dust, should thus increase the average dust temperature ($T_{\rm dust}$) of the system, so galaxies with a higher AGN contribution (parametrised as $\phi_{\rm IR, AGN}$/$\phi_{\rm IR}$) should have higher $T_{\rm dust}$. We found that $\phi_{\rm IR, AGN}$/$\phi_{\rm IR}$ is at $<3$ per cent until the LIRG regime, increasing rapidly thereafter, hence we would expect $T_{\rm dust}$ to follow the same trend as a function of $L_{\rm IR}$ with little or no change until the LIRG regime and a fast change thereafter. Indeed such a relation is observed in local galaxies (e.g. Dunne et al. 2000\nocite{Dunne00}; Dale et al. 2001\nocite{Dale01}; Dale $\&$ Helou 2002\nocite{DH02}; Chapman et al. 2003\nocite{Chapman03}; Chapin et al. 2009\nocite{CHA09}; S13): the average dust temperature of local galaxies is almost constant at $\sim$ 29-31\,K for $L_{\rm IR}<10^{11} L_{\odot}$, but undergoes a rapid increase in the LIRG and ULIRG regimes, amounting to a change of about 15\,K, mirroring  the change in the AGN contribution as a function of $L_{\rm IR}$ (Fig \ref{fig:fig3}). It is thus plausible that the local $L_{\rm IR}$--$T_{\rm dust}$ relation is driven by AGN and hence the high average dust temperatures seen in ULIRGs are a result of the increased AGN contribution to dust heating. 

Clues that this might indeed be the case, also come from earlier works which examined the relation between AGN signatures and SED shape. Several works showed that flatter mid-IR SEDs, i.e. warm mid-IR colours ($f_{\nu}$ (25$\mu$m)/$f_{\nu}$ (60$\mu$m)$>$0.2) indicate the presence of an AGN (e.g. de Grijp et al. 1985\nocite{deGrijp85}; 1987\nocite{deGrijp87}; 1992\nocite{deGrijp92}; Sanders et al. 1988b). Moreover, the combined findings of S90 and G11 indicate that for $L_{\rm IR}>10^{11}$~L$_{\odot}$ the local IR luminosity function is dominated by galaxies which host AGN (G11) and which have warm ($>$36\,K) average dust temperatures (S90).

\subsection{What powers local ULIRGs?}
\label{sec:power}

The principal outstanding question since the discovery of ULIRGs by the \textit{IRAS} all sky survey, is whether they are powered by AGN or star-formation. A conference in 1999 was solely devoted to exploring this question, culminating in the `Great Debate' (Sanders 1999; Joseph 1999). The `Great Debate' was originally aimed at evaluating the relative AGN/star formation contribution in individual ULIRGs and was later formulated as the question: `Do More than 50 per cent of local ULIRGs have more than 50 per cent of their emission powered by the AGN?'. A consensus was not reached and the answer was `Possibly' (Sanders 1999). Although our work does not address the AGN contribution in individual galaxies, we believe it provides a statistical answer to the aforementioned question. We find the answer is `No'. Under the assumption that the local Comton thick fraction is of the order of 30 per cent, we find that the luminosity above which ULIRGs become AGN-dominated (i.e. $\phi_{\rm IR, AGN}/\phi_{\rm IR}$ $>50\%$) falls within the range of log\,[$L_{\rm IR}/\rm L_{\odot}]\sim12.2$ to log\,[$L_{\rm IR}/\rm L_{\odot}] \sim12.7$ (Fig. \ref{fig:fig2}). As $\phi_{\rm IR}$ is a steeply declining function of $L_{\rm IR}$, the majority of ULIRGs have log\,$L_{\rm IR}/L_{\odot}<12.2$ (Fig. \ref{fig:fig1}), implying that the local ULIRG population is predominantly powered by star-formation: more than 50 per cent of local ULIRGs have more than 50 per cent of their emission powered by star-formation. However, if we ask the same question of the most luminous ULIRGs (log\,[$L_{\rm IR}/L_{\odot}]>12.7$), the reverse is true and they are predominantly powered by AGN.

\section{Conclusions}
\label{sec:conclusions}
We have examined the behaviour of the infrared galaxy luminosity function and the infrared AGN luminosity function in the local ($z<0.1$) Universe. The former corresponds to emission from dust heated by AGN and starlight, whereas the latter includes emission from AGN-heated dust only. We conclude that:
\begin{itemize}
\item The local infrared luminosity function is flatter at the high luminosity end than galaxy luminosity functions derived at other wavelengths such as the UV and optical, because of the increased AGN contribution to the galaxies' infrared emission with increasing luminosity. Infrared emission from AGN starts mixing into the galaxy luminosity function in the LIRG regime, constituting up to 30 per cent of the total infrared emission, and becomes significant in the ULIRG regime where it reaches 100 per cent. 
\item The local $L_{\rm IR}$-$T_{\rm dust}$ relation is plausibly driven by the increased AGN contribution to the galaxies' infrared emission with increasing infrared luminosity.
\item The local ULIRG population is primarily powered by star-formation: more than 50 per cent of local ULIRGs have more than 50 per cent of their emission powered by star-formation. The reverse is true for the most luminous (log\,[$L_{\rm IR}/L_{\odot}]>12.7$) ULIRGs however, and they are predominantly powered by AGN.
\end{itemize}

\bibliographystyle{mn2e}
\bibliography{references}

\begin{thebibliography}{}

\bibitem[\protect\citeauthoryear{{Armus} et~al.,}{{Armus}
  et~al.}{2006}]{Armus06}
{Armus} L.,  et~al., 2006, \apj, 640, 204

\bibitem[\protect\citeauthoryear{{Armus} et~al.,}{{Armus}
  et~al.}{2007}]{Armus07}
{Armus} L.,  et~al., 2007, \apj, 656, 148

\bibitem[\protect\citeauthoryear{{Brand}, {Dey}, {Weedman}, {Desai}, {Le
  Floc'h}, {Jannuzi}, {Soifer}, {Brown}, {Eisenhardt}, {Gorjian}, {Papovich},
  {Smith}, {Willner} \& {Cool}}{{Brand} et~al.}{2006}]{Brand06}
{Brand} K.,  {Dey} A.,  {Weedman} D.,  {Desai} V.,  {Le Floc'h} E.,  {Jannuzi}
  B.~T.,  {Soifer} B.~T.,  {Brown} M.~J.~I.,  {Eisenhardt} P.,  {Gorjian} V.,
  {Papovich} C.,  {Smith} H.~A.,  {Willner} S.~P.,    {Cool} R.~J.,  2006,
  \apj, 644, 143

\bibitem[\protect\citeauthoryear{{Brightman} \& {Nandra}}{{Brightman} \&
  {Nandra}}{2011}]{BN11b}
{Brightman} M.,  {Nandra} K.,  2011, \mnras, 414, 3084

\bibitem[\protect\citeauthoryear{{Chapin}, {Hughes} \& {Aretxaga}}{{Chapin}
  et~al.}{2009}]{CHA09}
{Chapin} E.~L.,  {Hughes} D.~H.,    {Aretxaga} I.,  2009, \mnras, 393, 653

\bibitem[\protect\citeauthoryear{{Chapman}, {Helou}, {Lewis} \&
  {Dale}}{{Chapman} et~al.}{2003}]{Chapman03}
{Chapman} S.~C.,  {Helou} G.,  {Lewis} G.~F.,    {Dale} D.~A.,  2003, \apj,
  588, 186

\bibitem[\protect\citeauthoryear{{Clements}, {Dunne} \& {Eales}}{{Clements}
  et~al.}{2010}]{CDE10}
{Clements} D.~L.,  {Dunne} L.,    {Eales} S.,  2010, \mnras, 403, 274

\bibitem[\protect\citeauthoryear{{Clements} et~al.,}{{Clements}
  et~al.}{2018}]{Clements18}
{Clements} D.~L.,  et~al., 2018, \mnras, 475, 2097

\bibitem[\protect\citeauthoryear{{Dale} \& {Helou}}{{Dale} \&
  {Helou}}{2002}]{DH02}
{Dale} D.~A.,  {Helou} G.,  2002, \apj, 576, 159

\bibitem[\protect\citeauthoryear{{Dale}, {Helou}, {Contursi}, {Silbermann} \&
  {Kolhatkar}}{{Dale} et~al.}{2001}]{Dale01}
{Dale} D.~A.,  {Helou} G.,  {Contursi} A.,  {Silbermann} N.~A.,    {Kolhatkar}
  S.,  2001, \apj, 549, 215

\bibitem[\protect\citeauthoryear{{Davies}, {Burston} \& {Ward}}{{Davies}
  et~al.}{2002}]{DBW02}
{Davies} R.~I.,  {Burston} A.,    {Ward} M.~J.,  2002, \mnras, 329, 367

\bibitem[\protect\citeauthoryear{{de Grijp}, {Keel}, {Miley}, {Goudfrooij} \&
  {Lub}}{{de Grijp} et~al.}{1992}]{deGrijp92}
{de Grijp} M.~H.~K.,  {Keel} W.~C.,  {Miley} G.~K.,  {Goudfrooij} P.,    {Lub}
  J.,  1992, \aaps, 96, 389

\bibitem[\protect\citeauthoryear{{de Grijp}, {Lub} \& {Miley}}{{de Grijp}
  et~al.}{1987}]{deGrijp87}
{de Grijp} M.~H.~K.,  {Lub} J.,    {Miley} G.~K.,  1987, \aaps, 70, 95

\bibitem[\protect\citeauthoryear{{de Grijp}, {Miley}, {Lub} \& {de Jong}}{{de
  Grijp} et~al.}{1985}]{deGrijp85}
{de Grijp} M.~H.~K.,  {Miley} G.~K.,  {Lub} J.,    {de Jong} T.,  1985, \nat,
  314, 240

\bibitem[\protect\citeauthoryear{{Dunne}, {Eales}, {Edmunds}, {Ivison},
  {Alexander} \& {Clements}}{{Dunne} et~al.}{2000}]{Dunne00}
{Dunne} L.,  {Eales} S.,  {Edmunds} M.,  {Ivison} R.,  {Alexander} P.,
  {Clements} D.~L.,  2000, \mnras, 315, 115

\bibitem[\protect\citeauthoryear{{Franceschini} et~al.,}{{Franceschini}
  et~al.}{2003a}]{Franceschini03b}
{Franceschini} A.,  et~al., 2003a, \mnras, 343, 1181

\bibitem[\protect\citeauthoryear{{Franceschini} et~al.,}{{Franceschini}
  et~al.}{2003b}]{Franceschini03}
{Franceschini} A.,  et~al., 2003b, \mnras, 343, 1181

\bibitem[\protect\citeauthoryear{{Genzel}, {Lutz}, {Sturm}, {Egami}, {Kunze},
  {Moorwood}, {Rigopoulou}, {Spoon}, {Sternberg}, {Tacconi-Garman}, {Tacconi}
  \& {Thatte}}{{Genzel} et~al.}{1998}]{Genzel98}
{Genzel} R.,  {Lutz} D.,  {Sturm} E.,  {Egami} E.,  {Kunze} D.,  {Moorwood}
  A.~F.~M.,  {Rigopoulou} D.,  {Spoon} H.~W.~W.,  {Sternberg} A.,
  {Tacconi-Garman} L.~E.,  {Tacconi} L.,    {Thatte} N.,  1998, \apj, 498, 579

\bibitem[\protect\citeauthoryear{{Goto} et~al.,}{{Goto}
  et~al.}{2011}]{Goto11a}
{Goto} T.,  et~al., 2011, \mnras, 410, 573

\bibitem[\protect\citeauthoryear{{Gregorich}, {Neugebauer}, {Soifer}, {Gunn} \&
  {Herter}}{{Gregorich} et~al.}{1995}]{Gregorich95}
{Gregorich} D.~T.,  {Neugebauer} G.,  {Soifer} B.~T.,  {Gunn} J.~E.,
  {Herter} T.~L.,  1995, \aj, 110, 259

\bibitem[\protect\citeauthoryear{{Guainazzi}, {Matt} \& {Perola}}{{Guainazzi}
  et~al.}{2005}]{Guainazzi05}
{Guainazzi} M.,  {Matt} G.,    {Perola} G.~C.,  2005, \aap, 444, 119

\bibitem[\protect\citeauthoryear{{Houck} et~al.,}{{Houck}
  et~al.}{1984}]{Houck84}
{Houck} J.~R.,  et~al., 1984, \apjl, 278, L63

\bibitem[\protect\citeauthoryear{{Houck}, {Schneider}, {Danielson},
  {Neugebauer}, {Soifer}, {Beichman} \& {Lonsdale}}{{Houck}
  et~al.}{1985}]{Houck85}
{Houck} J.~R.,  {Schneider} D.~P.,  {Danielson} G.~E.,  {Neugebauer} G.,
  {Soifer} B.~T.,  {Beichman} C.~A.,    {Lonsdale} C.~J.,  1985, \apjl, 290, L5

\bibitem[\protect\citeauthoryear{{Imanishi}, {Dudley}, {Maiolino}, {Maloney},
  {Nakagawa} \& {Risaliti}}{{Imanishi} et~al.}{2007}]{Imanishi07}
{Imanishi} M.,  {Dudley} C.~C.,  {Maiolino} R.,  {Maloney} P.~R.,  {Nakagawa}
  T.,    {Risaliti} G.,  2007, \apjs, 171, 72

\bibitem[\protect\citeauthoryear{{Imanishi}, {Maiolino} \&
  {Nakagawa}}{{Imanishi} et~al.}{2010}]{IMN10}
{Imanishi} M.,  {Maiolino} R.,    {Nakagawa} T.,  2010, \apj, 709, 801

\bibitem[\protect\citeauthoryear{{Imanishi}, {Nakagawa}, {Ohyama}, {Shirahata},
  {Wada}, {Onaka} \& {Oi}}{{Imanishi} et~al.}{2008}]{Imanishi08}
{Imanishi} M.,  {Nakagawa} T.,  {Ohyama} Y.,  {Shirahata} M.,  {Wada} T.,
  {Onaka} T.,    {Oi} N.,  2008, \pasj, 60, S489

\bibitem[\protect\citeauthoryear{{Iwasawa} et~al.,}{{Iwasawa}
  et~al.}{2011}]{Iwasawa11}
{Iwasawa} K.,  et~al., 2011, \aap, 529, A106

\bibitem[\protect\citeauthoryear{{Johnson}}{{Johnson}}{1966}]{Johnson66}
{Johnson} H.~L.,  1966, \apj, 143, 187

\bibitem[\protect\citeauthoryear{{Joseph}}{{Joseph}}{1999}]{Joseph99}
{Joseph} R.~D.,  1999, \apss, 266, 321

\bibitem[\protect\citeauthoryear{{Klaas}, {Haas}, {Heinrichsen} \&
  {Schulz}}{{Klaas} et~al.}{1997}]{Klaas97}
{Klaas} U.,  {Haas} M.,  {Heinrichsen} I.,    {Schulz} B.,  1997, \aap, 325,
  L21

\bibitem[\protect\citeauthoryear{{Klaas}, {Haas}, {M{\"u}ller}, {Chini},
  {Schulz}, {Coulson}, {Hippelein}, {Wilke}, {Albrecht} \& {Lemke}}{{Klaas}
  et~al.}{2001}]{Klaas01}
{Klaas} U.,  {Haas} M.,  {M{\"u}ller} S.~A.~H.,  {Chini} R.,  {Schulz} B.,
  {Coulson} I.,  {Hippelein} H.,  {Wilke} K.,  {Albrecht} M.,    {Lemke} D.,
  2001, \aap, 379, 823

\bibitem[\protect\citeauthoryear{{Kleinmann} \& {Low}}{{Kleinmann} \&
  {Low}}{1970}]{KL70}
{Kleinmann} D.~E.,  {Low} F.~J.,  1970, \apjl, 159, L165+

\bibitem[\protect\citeauthoryear{{Lawrence}, {Walker}, {Rowan-Robinson},
  {Leech} \& {Penston}}{{Lawrence} et~al.}{1986}]{Lawrence86}
{Lawrence} A.,  {Walker} D.,  {Rowan-Robinson} M.,  {Leech} K.~J.,    {Penston}
  M.~V.,  1986, \mnras, 219, 687

\bibitem[\protect\citeauthoryear{{Low} \& {Tucker}}{{Low} \&
  {Tucker}}{1968}]{LT68}
{Low} F.~J.,  {Tucker} W.~H.,  1968, Physical Review Letters, 21, 1538

\bibitem[\protect\citeauthoryear{{Maiolino}, {Salvati}, {Bassani}, {Dadina},
  {della Ceca}, {Matt}, {Risaliti} \& {Zamorani}}{{Maiolino}
  et~al.}{1998}]{Maiolino98}
{Maiolino} R.,  {Salvati} M.,  {Bassani} L.,  {Dadina} M.,  {della Ceca} R.,
  {Matt} G.,  {Risaliti} G.,    {Zamorani} G.,  1998, \aap, 338, 781

\bibitem[\protect\citeauthoryear{{Maiolino}, {Shemmer}, {Imanishi}, {Netzer},
  {Oliva}, {Lutz} \& {Sturm}}{{Maiolino} et~al.}{2007}]{Maiolino07}
{Maiolino} R.,  {Shemmer} O.,  {Imanishi} M.,  {Netzer} H.,  {Oliva} E.,
  {Lutz} D.,    {Sturm} E.,  2007, \aap, 468, 979

\bibitem[\protect\citeauthoryear{{Murakami} et~al.,}{{Murakami}
  et~al.}{2007}]{Murakami07}
{Murakami} H.,  et~al., 2007, \pasj, 59, 369

\bibitem[\protect\citeauthoryear{{Nandra} \& {Pounds}}{{Nandra} \&
  {Pounds}}{1994}]{NP94}
{Nandra} K.,  {Pounds} K.~A.,  1994, \mnras, 268, 405

\bibitem[\protect\citeauthoryear{{Nardini} \& {Risaliti}}{{Nardini} \&
  {Risaliti}}{2011}]{NR11}
{Nardini} E.,  {Risaliti} G.,  2011, \mnras, 415, 619

\bibitem[\protect\citeauthoryear{{Neugebauer} et~al.,}{{Neugebauer}
  et~al.}{1984}]{Neugebauer84}
{Neugebauer} G.,  et~al., 1984, \apjl, 278, L1

\bibitem[\protect\citeauthoryear{{Oyabu} et~al.,}{{Oyabu}
  et~al.}{2011}]{Oyabu11}
{Oyabu} S.,  et~al., 2011, \aap, 529, A122

\bibitem[\protect\citeauthoryear{{Page}, {Reeves}, {O'Brien} \&
  {Turner}}{{Page} et~al.}{2005}]{Page05}
{Page} K.~L.,  {Reeves} J.~N.,  {O'Brien} P.~T.,    {Turner} M.~J.~L.,  2005,
  \mnras, 364, 195

\bibitem[\protect\citeauthoryear{{Reeves} \& {Turner}}{{Reeves} \&
  {Turner}}{2000}]{RT00}
{Reeves} J.~N.,  {Turner} M.~J.~L.,  2000, \mnras, 316, 234

\bibitem[\protect\citeauthoryear{{Ricci}, {Ueda}, {Koss}, {Trakhtenbrot},
  {Bauer} \& {Gandhi}}{{Ricci} et~al.}{2015}]{Ricci15}
{Ricci} C.,  {Ueda} Y.,  {Koss} M.~J.,  {Trakhtenbrot} B.,  {Bauer} F.~E.,
  {Gandhi} P.,  2015, \apjl, 815, L13

\bibitem[\protect\citeauthoryear{{Rodr{\'{\i}}guez-Ardila} \&
  {Mazzalay}}{{Rodr{\'{\i}}guez-Ardila} \& {Mazzalay}}{2006}]{RAM06}
{Rodr{\'{\i}}guez-Ardila} A.,  {Mazzalay} X.,  2006, \mnras, 367, L57

\bibitem[\protect\citeauthoryear{{Sanders}}{{Sanders}}{1999}]{Sanders99}
{Sanders} D.~B.,  1999, \apss, 266, 331

\bibitem[\protect\citeauthoryear{{Sanders}, {Mazzarella}, {Kim}, {Surace} \&
  {Soifer}}{{Sanders} et~al.}{2003}]{Sanders03}
{Sanders} D.~B.,  {Mazzarella} J.~M.,  {Kim} D.-C.,  {Surace} J.~A.,
  {Soifer} B.~T.,  2003, \aj, 126, 1607

\bibitem[\protect\citeauthoryear{{Sanders} \& {Mirabel}}{{Sanders} \&
  {Mirabel}}{1996}]{SM96}
{Sanders} D.~B.,  {Mirabel} I.~F.,  1996, \araa, 34, 749

\bibitem[\protect\citeauthoryear{{Sanders}, {Phinney}, {Neugebauer}, {Soifer}
  \& {Matthews}}{{Sanders} et~al.}{1989}]{Sanders89}
{Sanders} D.~B.,  {Phinney} E.~S.,  {Neugebauer} G.,  {Soifer} B.~T.,
  {Matthews} K.,  1989, \apj, 347, 29

\bibitem[\protect\citeauthoryear{{Sanders}, {Soifer}, {Elias}, {Madore},
  {Matthews}, {Neugebauer} \& {Scoville}}{{Sanders} et~al.}{1988}]{Sanders88a}
{Sanders} D.~B.,  {Soifer} B.~T.,  {Elias} J.~H.,  {Madore} B.~F.,  {Matthews}
  K.,  {Neugebauer} G.,    {Scoville} N.~Z.,  1988, \apj, 325, 74

\bibitem[\protect\citeauthoryear{{Sanders}, {Soifer}, {Elias}, {Neugebauer} \&
  {Matthews}}{{Sanders} et~al.}{1988}]{Sanders88b}
{Sanders} D.~B.,  {Soifer} B.~T.,  {Elias} J.~H.,  {Neugebauer} G.,
  {Matthews} K.,  1988, \apjl, 328, L35

\bibitem[\protect\citeauthoryear{{Saunders}, {Rowan-Robinson}, {Lawrence},
  {Efstathiou}, {Kaiser}, {Ellis} \& {Frenk}}{{Saunders}
  et~al.}{1990}]{Saunders90}
{Saunders} W.,  {Rowan-Robinson} M.,  {Lawrence} A.,  {Efstathiou} G.,
  {Kaiser} N.,  {Ellis} R.~S.,    {Frenk} C.~S.,  1990, \mnras, 242, 318

\bibitem[\protect\citeauthoryear{{Sazonov}, {Revnivtsev}, {Krivonos},
  {Churazov} \& {Sunyaev}}{{Sazonov} et~al.}{2007}]{Sazonov07}
{Sazonov} S.,  {Revnivtsev} M.,  {Krivonos} R.,  {Churazov} E.,    {Sunyaev}
  R.,  2007, \aap, 462, 57

\bibitem[\protect\citeauthoryear{{Schechter}}{{Schechter}}{1976}]{Schechter76}
{Schechter} P.,  1976, \apj, 203, 297

\bibitem[\protect\citeauthoryear{{Soifer} et~al.,}{{Soifer}
  et~al.}{1984}]{Soifer84a}
{Soifer} B.~T.,  et~al., 1984, \apjl, 278, L71

\bibitem[\protect\citeauthoryear{{Soifer} et~al.,}{{Soifer}
  et~al.}{2000}]{Soifer00}
{Soifer} B.~T.,  et~al., 2000, \aj, 119, 509

\bibitem[\protect\citeauthoryear{{Soifer}, {Neugebauer}, {Helou}, {Lonsdale},
  {Hacking}, {Rice}, {Houck}, {Low} \& {Rowan-Robinson}}{{Soifer}
  et~al.}{1984}]{Soifer84b}
{Soifer} B.~T.,  {Neugebauer} G.,  {Helou} G.,  {Lonsdale} C.~J.,  {Hacking}
  P.,  {Rice} W.,  {Houck} J.~R.,  {Low} F.~J.,    {Rowan-Robinson} M.,  1984,
  \apjl, 283, L1

\bibitem[\protect\citeauthoryear{{Soifer}, {Sanders}, {Madore}, {Neugebauer},
  {Danielson}, {Elias}, {Lonsdale} \& {Rice}}{{Soifer} et~al.}{1987}]{Soifer87}
{Soifer} B.~T.,  {Sanders} D.~B.,  {Madore} B.~F.,  {Neugebauer} G.,
  {Danielson} G.~E.,  {Elias} J.~H.,  {Lonsdale} C.~J.,    {Rice} W.~L.,  1987,
  \apj, 320, 238

\bibitem[\protect\citeauthoryear{{Soifer}, {Sanders}, {Neugebauer},
  {Danielson}, {Lonsdale}, {Madore} \& {Persson}}{{Soifer}
  et~al.}{1986}]{Soifer86}
{Soifer} B.~T.,  {Sanders} D.~B.,  {Neugebauer} G.,  {Danielson} G.~E.,
  {Lonsdale} C.~J.,  {Madore} B.~F.,    {Persson} S.~E.,  1986, \apjl, 303, L41

\bibitem[\protect\citeauthoryear{{Symeonidis} et~al.,}{{Symeonidis}
  et~al.}{2013}]{Symeonidis13a}
{Symeonidis} M.,  et~al., 2013, \mnras, 431, 2317

\bibitem[\protect\citeauthoryear{{Symeonidis}, {Giblin}, {Page}, {Pearson},
  {Bendo}, {Seymour} \& {Oliver}}{{Symeonidis} et~al.}{2016}]{Symeonidis16}
{Symeonidis} M.,  {Giblin} B.~M.,  {Page} M.~J.,  {Pearson} C.,  {Bendo} G.,
  {Seymour} N.,    {Oliver} S.~J.,  2016, \mnras

\bibitem[\protect\citeauthoryear{{Symeonidis} \& {Page}}{{Symeonidis} \&
  {Page}}{2018}]{SP18}
{Symeonidis} M.,  {Page} M.~J.,  2018, \mnras, 479, L91

\bibitem[\protect\citeauthoryear{{Teng} et~al.,}{{Teng}  et~al.}{2009}]{Teng09}
{Teng} S.~H.,  et~al., 2009, \apj, 691, 261

\bibitem[\protect\citeauthoryear{{Tueller}, {Mushotzky}, {Barthelmy},
  {Cannizzo}, {Gehrels}, {Markwardt}, {Skinner} \& {Winter}}{{Tueller}
  et~al.}{2008}]{Tueller08}
{Tueller} J.,  {Mushotzky} R.~F.,  {Barthelmy} S.,  {Cannizzo} J.~K.,
  {Gehrels} N.,  {Markwardt} C.~B.,  {Skinner} G.~K.,    {Winter} L.~M.,  2008,
  \apj, 681, 113

\bibitem[\protect\citeauthoryear{{Vasudevan}, {Fabian}, {Gandhi}, {Winter} \&
  {Mushotzky}}{{Vasudevan} et~al.}{2010}]{Vasudevan10}
{Vasudevan} R.~V.,  {Fabian} A.~C.,  {Gandhi} P.,  {Winter} L.~M.,
  {Mushotzky} R.~F.,  2010, \mnras, 402, 1081

\end{thebibliography}

\end{document}